\documentclass[12pt,reqno]{elsart}

\usepackage{amsmath,amsfonts,graphicx}

\theoremstyle{remark}
\theoremstyle{remark}
\newtheorem{remark}{Remark}

\def\semip{\times}
\def\rd{\mathrm d}
\def\ri{\mathrm i}
\def\A{\mathcal A}
\def\C{\mathbb C}
\def\RP{\mathbb {RP}}
\def\R{\mathbb R}
\def\rd{\mathrm{d}}
\def\sl{\mathfrak{sl}}
\def\g{\mathfrak{g}}
\def\calS{\mathcal S}

\def\G{\mathcal{G}}

\def\SL{\mathrm{SL}}

\def\Lo{\mathcal{L}}
\def\Hop{\mathcal{H}}
\def\G{\mathcal{G}}

\def\Ad{\operatorname{Ad}}
\def\tr{\operatorname{trace}}
\def\sl{\mathfrak{sl}}
\def\n{\mathfrak{n}}

\def\Lo{\mathcal{L}}

\begin{document}
\runauthor{Calini, Ivey and Mar\'i-Beffa}
\begin{frontmatter}
\title{Remarks on KdV-type Flows on Star-Shaped Curves}
\author[CofC]{Annalisa Calini\thanksref{Someone}}
\author[CofC]{Thomas Ivey}
\author[UWM]{Gloria Mar\'i-Beffa}

\address[CofC]{Department of Mathematics, College of Charleston, Charleston SC 29424, USA}
\address[UWM]{Department of Mathematics, University of Wisconsin-Madison, Madison WI 53706, USA}
\thanks[Someone]{Corresponding author; e-mail: {\tt calinia@cofc.edu}}
\begin{abstract}
We study the relation between the centro-affine geometry of star-shaped planar curves 
and the projective geometry of parametrized maps into $\RP^1$. We show that projectivization induces a map between differential invariants and a bi-Poisson map between Hamiltonian structures.  We also show that a Hamiltonian evolution equation for closed star-shaped planar curves, discovered by Pinkall, has the Schwarzian KdV equation as its projectivization.  (For both flows, the curvature
evolves by the KdV equation.) Using algebro-geometric methods and the relation of group-based moving frames to AKNS-type representations, we construct examples of closed solutions of Pinkall's flow associated with periodic finite-gap KdV potentials.
\end{abstract}
\end{frontmatter}

\section{Introduction}
In his 1995 paper \cite{P95} U. Pinkall derived a Hamiltonian evolution equation on the space of closed star-shaped curves in the centro-affine plane which is closely related to the KdV equation.  (As explained in \S\ref{pinkallsection},
a planar curve is star-shaped if its position and velocity vectors are always
linearly independent; unless otherwise noted, we will restrict our attention
to planar curves satisfying this condition.)
The dynamics of curves in the centro-affine plane, as related to integrable hierarchies including KdV, 
were further explored in more recent work by K-S. Chou and C. Qu \cite{CQ1, CQ2}, 
in which concrete examples of traveling wave solutions of Pinkall's flow were presented. Meanwhile, for curves in $\RP^{n-1}$, evolution equations which are invariant under the $\mathrm{PSL}(n,\R)$ action have been shown 
(by the third author and her collaborators) to be Hamiltonian with respect to a Poisson structure related to the KdV hierarchy  \cite{LHM, M3}. In particular, the Schwarzian KdV equation is a $\mathrm{PSL}(2,\R)$-invariant evolution of parametrized curves in $\RP^1$ which is
closely related to the standard KdV equation.

In this article we begin a study of the connection between the projective and centro-affine geometry of curves, and of the relation between invariant curve evolutions in both geometries and associated geometric Hamiltonian structures. Geometric Hamiltonian structures are Poisson brackets defined on the space of differential invariants (or curvatures) of the curves (see \cite{M1}, \cite{M2}). Such Poisson brackets are known for many of the classical geometries, including projective, but not for the centro-affine case. 
For the projective case, 
two geometric Poisson brackets on the space of curvature functions
are given by the compatible Hamiltonian structures for the 
KdV equation, seen as a bi-Hamiltonian system.

In a similar way,
 we will exhibit two geometric Poisson brackets on the space of differential invariants
for curves in the centro-affine plane. 
We show that these brackets can be restricted to the submanifold where the speed is unity 
(i.e., we specialize to curves parametrized by centro-affine arclength),
and that the resulting restricted Poisson brackets again coincide with the 
first two canonical Poisson brackets of the KdV equation. Furthermore, 
we show that projectivization is a 1-to-1 map from differential
invariants of centro-affine curves parametrized by arc-length to the
differential invariants of parametrized curves in $\RP^1$.
(A posteriori, the curve parameter in the projective setting turns coincides with the centro-affine arclength.)
We also show that projectivization is a bi-Poisson map,
and that it takes an arc-length preserving evolution of 
planar curves, invariant under centro-affine transformations, 
to a $\mathrm{PSL}(2,\R)$-invariant evolution of curves in $\RP^1$. In particular, 
it takes solutions of Pinkall's flow to solutions of the Schwarzian KdV. 

One of the main tools of this part of the study is the use of group-based moving frames 
and the normalization procedure that produces them (see \cite{FO}). 
The relation of group-based moving frames to the AKNS representation of KdV 
is used in the last part of the paper to produce solutions of Pinkall's flow.
In this sense, we recover Pinkall's flow in two ways: 
as the Schwarzian KdV expressed in homogeneous coordinates, 
and as the evolution equation for a planar curve constructed 
in terms of components of a fundamental matrix solution of the AKNS system.

The usual phase space for Pinkall's flow is the space of closed unparametrized star-shaped curves.  (In fact, the evolution equation arises as a Hamiltonian flow on this space, with Hamiltonian functional given by the total area swept by the planar curve.)  The reconstruction formula for the centro-affine curve in terms of eigenfunctions of the AKNS system allow us to construct closed solutions of Pinkall's flow associated with periodic finite-gap KdV potentials.  (These are generalizations of periodic traveling wave solutions,
but with multiple phases.) 
The closure condition for such curves takes on a particularly simple form: given a periodic solution of the KdV equation (a periodic centro-affine curvature function), the associated star-shaped curve is closed, provided the ``normalization" parameter of the moving frame corresponds to a periodic point of the spectrum of the given KdV potential.

In order to construct explicit examples of such solutions, we bring together techniques and ideas previously used by the first two authors in the context of the vortex filament equation \cite{CI, CI2}, a geometric evolution equation associated with the integrable focusing nonlinear Schr\"odinger equation.  In particular, we combine the algebro-geometric construction of the Baker-Akhiezer eigenfunction of the KdV Lax pair \cite{bbeim} with the curve reconstruction formula from solutions of the AKNS system,  to produce expressions for general finite-gap solutions of Pinkall's flow. 

Implementation of the closure condition and construction of concrete examples of closed finite-gap solutions are most easily achieved by adapting the theory of isoperiodic deformations \cite{GS} to the KdV setting. In the examples appearing in last section of this paper, 
we deform the spectrum of the zero KdV potential, increasing the genus
 while preserving both periodicity and closure of the associated curve. In this way, we generate interesting examples of closed solutions of Pinkall's flow with an arbitrary number of linear phases.
As noted above, the
projectivization of these solutions are solutions of the Schwarzian KdV equation, but
properties of these solutions are better observed (as shown by our pictures) on the centro-affine plane. 

The study of higher dimensional cases is under way and will relate 
generalizations of the Schwarzian-KdV equation to generalizations of Pinkall's evolution, both of them linked to Adler-Gelfand-Dikii brackets and generalized KdV hierarchies (as defined in \cite{DS}).

\section{Pinkall's Flow for Star-Shaped Curves in $\R^2$}\label{pinkallsection}
From a group-based point of view, the centro-affine geometry of the plane is defined by
the linear action of $G={\rm SL}(2,\R)$ on $\R^2$.
Let $\gamma:I \to \R^2$ be a smooth curve parametrized by $x$,  where $I\subset \R$ is an open interval.
(We will occasionally specialize to the case where $\gamma$ is periodic.)
The curve is {\em star-shaped} if
\begin{equation}
\label{kepler}
\det(\gamma(x), \gamma'(x)) \ne 0 \quad \forall x,
\end{equation}
where prime denotes differentiation by $x$; this condition is clearly $G$-invariant.
The basic integral invariant of star-shaped curves is {\em centro-affine arclength}, given by
$$s = \int \det(\gamma, \gamma') \,\rd x.$$
Accordingly, we define $v=\det(\gamma, \gamma')$ to be the {\em centro-affine speed} of
$\gamma$.

If $\gamma$ is parametrized by arclength, then $\det\left(\gamma, \gamma_s\right)=1$. Differentiating
this with respect to $s$ gives the relation
\[
\gamma_{ss} = -p \gamma
\]
for some function $p=\det(\gamma_s, \gamma_{ss}) $, called the \emph{centro-affine curvature}.
(The formula for $p$ in terms of $x$ is more complicated.)
Centro-affine speed and curvature
are independent and generating \emph{differential invariants} of parametrized planar curves under
the linear action of $G$.

Pinkall's flow may be derived from a natural Hamiltonian structure on $M$, the space of closed, unparametrized star-shaped curves in $\R^2$.
An arclength-preserving vector field on $M$ has the form
$\displaystyle X=\frac{\beta_s }{2}\gamma-\beta \gamma_s$, for some smooth periodic function $\beta(s)$. Hamiltonian vector fields on $M$ are generated by means of the canonical symplectic form
\[
\omega(X,Y)=\oint \det(X,Y) \rd s
\]
which defines the following correspondence between a Hamiltonian functional $H$ and a Hamiltonian vector field $X_H$:
\[
\rd H(X)=\omega(X, X_H),  \quad \forall X\in T_\gamma M.
\]

In particular, choosing as in \cite{P95} the Hamiltonian $H=\oint p(s) \, \rd s$ gives Pinkall's flow
\begin{equation}\label{pinkall}
\gamma_t = \frac{p_s}{2} \gamma - p \gamma_s.
\end{equation}
In turn, under this flow the centro-affine curvature $p$ evolves by\[
p_t = -\tfrac12 p_{sss} - 3p p_s,
\]
a version of the KdV equation.

\section{Projective Geometry of Curves in $\RP^1$}

In this section we will briefly review the geometry of regular parametrized curves in $\RP^1$, their
group-based moving frames and their invariant flows.  (For more information on group-based
moving frames and the normalization equations that can be used to find them, see the original
paper \cite{FO} where they were introduced; see also \cite{M2} for the projective and other cases.)
Later, we define the Schwarzian flow for these curves, and 
relate invariant flows in this geometry to invariant flows for
centro-affine curves.

\subsection{Group-based Moving Frames}

From a group-based viewpoint, the geometry of $\RP^1$ is defined by
the action of $G=SL(2,\R)$ via linear fractional
transformations
$$\begin{pmatrix}a & b \\ c & d \end{pmatrix}
\cdot u = \dfrac{a u + b}{c u + d},
$$
(Here, $u$ is an affine coordinate on $\RP^1$.)

Let $\phi:I \to \RP^1$ be a regular map.  We will think of $u(x)$ as
giving the value of $\phi$ in the affine coordinate; without
loss of generality, we may assume that $u'(x)>0$.
We define an equivariant lift $\rho: I \to G$ of $\phi$, depending
on an arbitrary parameter $\lambda \in \R$, as follows.
For each $x\in I$, let $g(x) \in G$ be the group element
satisfying the following normalization conditions:
\begin{align*}
(g\cdot u)(x) &= 0,\\
(g \cdot u)'(x) &= 1,\\
(g \cdot u)''(x) &= 2\lambda,
\end{align*}
where  $\displaystyle (g \cdot u)'(x):=\left. \tfrac{\rd}{\rd y} g(x)\cdot u(y)\right|_{y=x}.$
One computes the following factorized form for the group element $g$:
$$g(x) =
\begin{pmatrix}1 & 0 \\ \dfrac{u''(x)}{2u'(x)} - \lambda & 1\end{pmatrix}
\begin{pmatrix}u'(x)^{-1/2} & 0 \\ 0 & u'(x)^{1/2}\end{pmatrix}
\begin{pmatrix}1 & -u(x) \\ 0 & 1 \end{pmatrix}.
$$
This element of the group defines a \emph{right moving frame} for $u$, as defined in \cite{FO}. 
In order to obtain a lift to $G$ that is equivariant for the {\em left action}
of $G$ on itself (a left moving frame, \cite{FO}), we choose its inverse
$$\rho(x) = g(x)^{-1} =
\begin{pmatrix}1 & u(x) \\ 0 & 1 \end{pmatrix}
\begin{pmatrix}u'(x)^{1/2} & 0 \\ 0 & u'(x)^{-1/2}\end{pmatrix}
\begin{pmatrix}1 & 0 \\ \lambda-\dfrac{u''(x)}{2u'(x)}& 1\end{pmatrix},
$$
%By analogy with the construction of Euclidean invariants for curves through
%taking the derivative of the Frenet frame,
and refer to $\rho$ as a {\em
$\lambda$-normalized left moving frame} for $\phi:\R \to \RP^1$.
%When $\lambda=0$, we will call $\rho$ the
%{\em standard normalized (equivariant) projective frame} for $\phi$.

The matrix-valued function $\rho^{-1} \rho'$ is called the \emph{Maurer-Cartan matrix} associated
to $\rho$. The entries of this matrix contain an independent,  generating set
of differential invariants for the curve, i.e., any other differential invariant
is a function of  the entries of $\rho^{-1}\rho'$ and their derivatives (see \cite{H} for the general result).

\begin{prop}
The $\lambda$-normalized  left moving frame $\rho$ satisfies the differential equation
\begin{equation}\label{rhoxform}
\rho^{-1} \rho' = \begin{pmatrix} \lambda & 1 \\ \kappa - \lambda^2 & -\lambda \end{pmatrix},
\end{equation}
where $\kappa = -\tfrac12 \calS(u)$
is the {\em projective curvature} of $\phi$, and
\[
\calS(f) = \dfrac{f'''}{f'} - \frac32 \left(\dfrac{f''}{f'}\right)^2
\]
is the Schwarzian derivative.
\end{prop}

\begin{remark}
The calculation of \eqref{rhoxform} is simplified by using the factorization
\begin{equation}\label{groupfactor}
G = G_{-1} \cdot G_0 \cdot G_1
\end{equation}
for elements of $G=SL(2,\R)$, where
$$
G_0 = \left\{ \begin{pmatrix}  a & 0 \\ 0 & a^{-1}\end{pmatrix}, a \in \R^*\right\},
\quad
G_1 = \left\{ \begin{pmatrix} 1 & 0 \\ b & 1\end{pmatrix}, b\in \R\right\}
\quad
G_{-1} = \left\{ \begin{pmatrix}1 & c \\ 0 & 1\end{pmatrix}, c \in \R \right\}.
\quad
$$
This mirrors the Lie algebra decomposition
\begin{equation}%\label{algebrasum}
\sl(2,\R) = \g_{-1} \oplus \g_0 \oplus \g_1,
\end{equation}
where
$$
\g_0 = \left\{ \begin{pmatrix}  \alpha & 0 \\ 0 & -\alpha\end{pmatrix}\right\},
\quad
\g_1 = \left\{\begin{pmatrix} 0 &0 \\ \beta & 0 \end{pmatrix} \right\},
\quad
\g_{-1} = \left\{\begin{pmatrix} 0&\gamma \\ 0 &0\end{pmatrix}\right\}.
$$
\end{remark}

\subsection{Invariant projective flows}
We now discuss how the  projective curvature $\kappa$ and the normalized projective frame $\rho$ evolve when
the curve $\phi$  satisfies an evolution equation, invariant under the $SL(2, \R)$ action on $\RP^1$.  
In affine coordinates, the most general such flow takes the form
\begin{equation}\label{utrform}
u_t = r u',
\end{equation}
where $r$ is a function of $\kappa$ and its derivatives \cite{M3}.
We will later specialize to the case $r=\kappa$, and show that in this case
the curvature and frame satisfy the KdV equation and its AKNS system, respectively.

\begin{prop}\label{timederivatives}
Suppose that $\phi$ evolves by \eqref{utrform}.  Then the $\lambda$-normalized
projective frame of $\phi$ satisfies
\begin{equation}\label{rhotform}
\rho^{-1}\rho_t = \begin{pmatrix} \lambda r + \tfrac12 r' & r \\
-\tfrac12 r'' -\lambda r' + (\kappa - \lambda^2) r  & -\lambda r - \tfrac12 r'
\end{pmatrix},
\end{equation}
and its projective curvature satisfies
\begin{equation}\label{kappatform}
\kappa_t = -\tfrac12 r''' + 2 \kappa r' + r\kappa'.
\end{equation}
\end{prop}
\begin{pf}
The computation is simplified if one uses the decomposition \eqref{groupfactor} to
write $\rho = \rho_{-1} \rho_0 \rho_{1}$, and
\begin{equation}\label{awfult}
\rho^{-1}\rho_t = \Ad_{\rho_1^{-1} \rho_0^{-1}}
\left( \rho_{-1}^{-1} (\rho_{-1})_t \right)
+ \Ad_{\rho_1^{-1} }\left( \rho_0^{-1} (\rho_0)_t \right)
+ \rho_1^{-1} (\rho_1)_t.
\end{equation}
Introducing the variable $w$ defined by $w= -\tfrac12 \log u'$, 
we compute $\displaystyle w'=-\frac{u''}{2u'};$ in addition, 
rewriting equation \eqref{utrform} as $u_t=re^{-2w}$, we obtain
\begin{align*}
\rho^{-1}\rho_t
%&= \Ad\left(
%\begin{pmatrix}1 & 0 \\ -\lambda - w' & 1\end{pmatrix}
% \begin{pmatrix}e^{w} & 0 \\ 0 & e^{-w}\end{pmatrix}
% \right)
%\left\{
%\begin{pmatrix}1 & -\gamma \\ 0 & 1 \end{pmatrix}
%\begin{pmatrix} 0 & \gamma_t \\ 0 & 0 \end{pmatrix}
% \right\} \\
% &+ \Ad \begin{pmatrix}1 & 0 \\ -\lambda - w' & 1\end{pmatrix}
% \left\{ \begin{pmatrix} -w_t & 0 \\ 0 & w_t \end{pmatrix} \right\}
% +\begin{pmatrix}1 & 0 \\ -\lambda - w' & 1\end{pmatrix}
% \begin{pmatrix} 0 & 0 \\ w_{xt} & 0 \end{pmatrix}
% \\
% &= \Ad \begin{pmatrix}1 & 0 \\ -\lambda - w' & 1\end{pmatrix}
% \left\{ \begin{pmatrix} -w_t & r \\ 0 & w_t \end{pmatrix} \right\}
% + \begin{pmatrix} 0 & 0 \\ w_{xt} & 0 \end{pmatrix} \\
 &= \begin{pmatrix} r(\lambda+w') - w_t & r \\ w'_t+
 (2w_t - r(\lambda+w'))(\lambda+w') & -r(\lambda+w') +w_t \end{pmatrix}.
\end{align*}
The time derivative of $w$ is readily computed from
$$w_t = -\dfrac12 \dfrac{u'_{t}}{u'}
= -\dfrac12 \dfrac{(ru')'}{u'} = -\dfrac12 r' + r w',$$
showing that the (1,1)-entry of $\rho^{-1}\rho_t$ is
$\lambda r + \tfrac12 r'$.  From
\begin{equation}\label{wxt}
w'_{t} = -\tfrac12 r'' + r' w' + r w'',
\end{equation}
there follows that the (2,1)-entry of $\rho^{-1}\rho_t$ is
$$-\tfrac12 r'' - \lambda r' + (\kappa - \lambda^2) r.$$

Finally, differentiating $\kappa = w'' + (w')^2$ with respect to $t$ and using equation \eqref{wxt} gives
\eqref{kappatform}.
%\begin{align*}\kappa_t &= w_{xxt} + 2w' w_{xt} =
%-\tfrac12 r''' + r'' w' + 2r' w'' + (\kappa - (w')^2)' r
%+ 2 w' (-\tfrac12 r'' + r' w' + r w'')\\
%&= -\frac12 r''' + 2\kappa r' + r \kappa'.
%\end{align*}
\qed
\end{pf}

%\begin{remark}
%The decomposition \eqref{algebrasum} is a gradation.  In particular,
%because $[ \g_0, \g_1] \subset \g_1$, only the first term in \eqref{awfult}
%can contain any component in $\g_{-1}$.  Moreover, because
%$[\g_1,\g_{-1}]\subset \g_0$, applying $\Ad_{\rho_1^{-1}}$
%to this term does not change the $\g_{-1}$ component.  Thus, the $\g_{-1}$ term
%in $\rho^{-1} \rho_t$ is simply
%$$\Ad_{\rho_0^{-1}}\left( \rho_{-1}^{-1} (\rho_{-1})_t \right)
%= \begin{pmatrix} 0 & u_t/u' \\ 0 & 0 \end{pmatrix}=\begin{pmatrix} 0 & r \\ 0 & 0 \end{pmatrix}.$$
%
%Furthermore, with an extra assumption,  we can obtain the
%rest of Proposition \ref{timederivatives} by the following alternate route.  Let
%\begin{equation}\label{rhoNform}
%\rho^{-1}\rho_t = N = \begin{pmatrix} \alpha & r \\ \beta & -\alpha \end{pmatrix}
%\end{equation}
%and  $\rho^{-1}\rho' = K,$ given by the right-hand side of \eqref{rhoxform}.  We assume
%that $N$ satisfies the compatibility condition
%\begin{equation}\label{KNcompatible}
%K_t  - N' - [K,N]=0,
%\end{equation}
%required for  the  system of equations \eqref{rhoxform} and \eqref{rhoNform}
%to be solvable.  Then, writing  equation \eqref{KNcompatible} component-wise gives both \eqref{rhotform} and \eqref{kappatform}.
%\end{remark}

Notice that, when $\displaystyle r=\kappa=-\tfrac12 \calS(u)$ in equation \eqref{utrform}, 
then equation \eqref{kappatform} reduces to a KdV equation
$$\kappa_t=-\tfrac12\kappa'''+3\kappa\kappa'.$$
It is a well-known fact \cite{KN, We} that the equation
\begin{equation}\label{SKdV}
u_t = -\tfrac12 \calS(u) u'
\end{equation}
 induces a KdV evolution for  $\displaystyle \tfrac12 \calS(u)$.
Indeed, equation \eqref{SKdV} is traditionally called the {\it Schwarzian KdV equation} because of this property.

\subsection{Relation between invariant evolutions}\label{relatethem}
%\section{Pinkall's flow and the Schwarzian KdV equation}
We conclude this section with a discussion of the relation between
invariant centro-affine and projective flows for curves---in particular,
between Pinkall's flow and the Schwarzian KdV equation.

Composition of a star-shaped curve $\gamma:I \to \R^2$ with
the projectivization map $\pi :\R^2 \to \RP^1$ gives a regular  map
$\phi:I \to\RP^1$, since $\gamma(x) \ne 0$ and the
 tangent of $\gamma$ is never zero or in the radial direction.
 Let $\gamma = (\gamma_1, \gamma_2)^T$ and assume that $\gamma_1 \ne 0$ on
an open interval.  Then, in an affine coordinate chart, the projectivization $\phi$
of $\gamma$ is given by
\[
u = \frac{\gamma_2}{\gamma_1}.\]
(Of course,
one would use $u = \gamma_1/\gamma_2$ near points where $\gamma_1$ vanishes.)

The most general flow for parametrized planar curves $\gamma(x)$ that is invariant under the centro-affine action is of the form
\[
\gamma_t %=\rho \begin{pmatrix} r_1\\ r_2\end{pmatrix}
= r_1 \gamma + \frac1{v} r_2 \gamma_x
\]
where  $r_1$ and $r_2$ are arbitrary differential invariants
and $v$ is the centro-affine speed.  If we begin with a curve parametrized by centro-affine arclength $s$,
then in order for unit speed to be preserved, we need $2r_1 + \frac{d}{ds} r_2 = 0$,
so that
\begin{equation}\label{ievca}
\gamma_t = -\frac12 r_s \gamma + r \gamma_s,
\end{equation}
for a differential invariant $r$.

The following proposition shows that $\pi$ takes centro-affine invariant evolutions to projective invariant evolutions  and that Pinkall's flow \eqref{pinkall} is simply the Schwarzian KdV equation in homogeneous coordinates.  (Note, however, that under projectivization the curvature $\kappa$ corresponds to $-1$ times Pinkall's curvature $p$.)

\begin{prop} Let $\gamma(x)$ be a star-shaped curve parametrized
by centro-affine arclength, and let $u(x)$ be its projectivization in an affine chart.
Then the centro-affine curvature satisfies $p(x) = \frac12 \calS(u)$.
 Furthermore, if $\gamma$ evolves by \eqref{ievca} then $u={\gamma_2}/{\gamma_1}$ evolves by
\[
u_t = r u'.
\]
In particular, if   $\gamma$ evolves by \eqref{pinkall} , then $u$ evolves by the Schwarzian KdV equation.
\end{prop}

\begin{pf}
The normalization condition $\det(\gamma, \gamma') = 1$ implies that
\[
1 = \gamma_1^2 \det\begin{pmatrix}1&\frac{\gamma_1'}{\gamma_1}\\ \frac{\gamma_2}{\gamma_1}&\frac{\gamma_2'}{\gamma_1}\end{pmatrix} 
= \gamma_1^2 \left( \frac{\gamma_2'}{\gamma_1} -\frac{\gamma_1'\gamma_2}{\gamma_1^2} \right)
=\gamma_1^2 u',
\]
Replacing  $\gamma_1$ by $(u')^{-\frac12}$and $\gamma_2$ by $u\gamma_1$ in the formula
\begin{equation}\label{peqn}
p = \det\begin{pmatrix} \gamma_1'&\gamma_1''\\ \gamma_2'&\gamma_2''\end{pmatrix},
\end{equation}
one obtains that $p = \frac12 \calS(u)$.
Finally, if $\displaystyle \gamma_t = -\tfrac{r_2'}2 \gamma + r_2\gamma'$, then
\[
u_t = \frac{(\gamma_2)_t}{\gamma_1} - \frac{\gamma_2}{\gamma_1}\frac{(\gamma_1)_t}{\gamma_1}
=
\frac{ -\frac{r_2'}2 \gamma_2 + r_2\gamma_2'}{\gamma_1}
- \frac{\gamma_2}{\gamma_1}\left(- \frac{r_2'}2  + r_2\frac{\gamma_1'}{\gamma_1}\right) 
= r_2 u'.
\]
\qed
\end{pf}

\section{Geometric Hamiltonian Structures}
In this section we calculate group-based moving frames and geometric Hamiltonian
structures, first for parametrized curves in $\RP^1$, and then
for curves in the centro-affine plane.  In each case,
we can restrict a pair of Poisson brackets on a loop algebra
to the space of curvature functions, and this results in
the first and second Poisson operators of the KdV equation. 
We also note that the projectivization
map $\pi:\R^2 \to \RP^1$ induces a bi-Poisson isomorphism between the 
spaces of differential invariants.

\subsection{Projective-geometric Hamiltonian structures}
In this section we restrict our attention to closed curves $\phi: S^1 \to \RP^1$.
(In other words, we take $I=\R$, the whole line, and assume that $\phi$ is $2\pi$-periodic.)
Thus, the Maurer-Cartan matrix \eqref{rhoxform} is periodic, and gives an element of
the loop algebra $\Lo\g = C^\infty(S^1,\g)$ where $\g=\sl(2,\R)$.

The space of projective curvature functions has a natural bi-Hamiltonian structure, 
obtained via reduction from a well-known bi-Hamiltonian structure on $\Lo\g^\ast$, where $\g^\ast$ is the dual of the Lie algebra $\g$. 
(We identify $\sl(2)$ with $\sl(2)^\ast$ in the standard fashion, using the Killing form.) 
Its definition is based on the following two results, which are specializations 
of more general theorems appearing in \cite{M2}.  
(The loop algebra, and the corresponding loop group $\Lo SL(2,\R)$, 
are endowed with the $C^\infty$ topology.)

\begin{prop}\label{projquot}
The space of periodic Maurer-Cartan matrices of the form \eqref{rhoxform} with $\lambda=0$, in
the neighborhood of a generic point, can be identified with the quotient
$\mathcal{M}/\Lo N$, where
$\mathcal{M} \subset \Lo\sl(2)^\ast$ is the submanifold consisting of loops of the form
\[
\begin{pmatrix} \alpha & 1\\ \beta & -\alpha\end{pmatrix},
\]
$N \subset SL(2,\R)$ is the subgroup of matrices of the form 
$\left(\begin{smallmatrix} 1 & 0 \\ * & 1 \end{smallmatrix}\right)$,
and $\Lo N$ acts on $\Lo\sl(2)^*$ via the gauge action
\begin{equation}\label{gaugeaction}
g \cdot L = g^{-1}g_x +g^{-1} L g, \qquad g\in \Lo N, L \in \Lo \sl(2)^*.
\end{equation}
\end{prop}

In the second result,  $\G,\Hop$ are functionals on $\Lo\g^\ast$,
and $\delta \G/\delta L, \delta \Hop/\delta L$ are their gradients at the point  $L\in \Lo \g^\ast$, which lie in $\Lo \g$.

\begin{thm}\label{projbr} The following compatible Poisson structures on $\Lo \g^\ast$, defined
by
\begin{align}\label{br1}
\{\Hop, \G\}(L)&= \int_{S^1}\tr\left(\left(
\left(\frac{\delta \Hop}{\delta L}\right)_x+
\left[L,\frac{\delta \Hop}{\delta L}\right]
\right) \frac{\delta \G}{\delta L}\right) dx, \\
\label{br2}
\{\Hop, \G\}_0(L) &= \int_{S^1}\tr\left(\left[\begin{pmatrix} 0&0\\ 1&0\end{pmatrix},
\frac{\delta \Hop}{\delta L}\right] \frac{\delta \G}{\delta L}\right) dx,
\end{align}
can be reduced to the quotient $\mathcal{M}/\Lo N$, to produce the second and first
 Hamiltonian structures, respectively, for KdV.
\end{thm}
%(Note that, by using the non-degenerate
%pairing between $\Lo \g$ and $\Lo \g^*$ given by
%integration over the circle, the variational derivatives $\dfrac{\delta \G}{\delta L}, \dfrac{\delta H}{\delta L}$ correspond to elements of $\Lo g$.)

The choice of the constant matrix 
$\left(\begin{smallmatrix} 0&0\\ 1&0\end{smallmatrix}\right)$ 
is specific for the projective case; different constant matrices are chosen for different geometries \cite{DS,M1}.
For an indication as to how the last two results are proved, see section \ref{camfs},
where the analogous results are proved for the centro-affine case.

The reduced brackets are examples of the so-called geometric Poisson brackets, Hamiltonian structures that are defined by the background geometry of the flow. Their derivation can be found in \cite{M2}.  We give their expressions as follows:
if $h(\kappa)$ and $f(\kappa)$ are two  functionals defined on the space of periodic curvatures for projective curves, the reduction of (\ref{br1}) is given by
\[
\{h,f\}(\kappa) =\int_{S^1} \frac{\delta f}{\delta \kappa} \left(-\frac12 D^3- \kappa D - D \kappa\right)  \frac{\delta h}{\delta \kappa} dx,
\]
where $D= \frac d{dx}$, while the reduction of (\ref{br2}) gives
\[
\{h,f\}_0(\kappa) = 2\int_{S^1} \frac{\delta f}{\delta \kappa} D   \frac{\delta h}{\delta \kappa} dx.
\]
Note that these brackets have been computed assuming that $\lambda=0$ in 
the moving frame; in general, the brackets will depend on the choice of $\lambda$.
%We will examine the role of non-zero $\lambda$ later in the paper.

\subsection{Centro-affine moving frames and geometric Hamiltonian structures} 
\label{camfs}
In this section we calculate group-based moving frames and geometric Hamiltonian
structures in a $C^\infty$ neighborhood of a generic centro-affine planare curve. 
We show that the Hamiltonian
structures can be restricted to the submanifold of constant centro-affine arc-length.
Like in the projective case, the restrictions are the first and second Poisson operators of the KdV equation. 
Finally, we show that the projectivization
map from the plane to $\RP^1$ induces a bi-Poisson isomorphism between the spaces of
differential invariants. 

\begin{prop} \label{camf} Let $\gamma:I \to \R^2$ be a star-shaped planar curve. The matrix
\[
\rho = \begin{pmatrix} \gamma &\frac 1v \gamma'\end{pmatrix}
\]
is a centro-affine left moving frame, where $v = \det(\gamma \hskip 2ex\gamma')$ 
is the centro-affine  speed. Its Maurer-Cartan matrix is given by
\begin{equation}\label{MC}
\rho^{-1} \rho_x = \begin{pmatrix} 0& -v p\\ v& 0\end{pmatrix},
\end{equation}
where $p$ is the centro-affine curvature of  $\gamma$. 
\end{prop}
\begin{pf} As in the projective case, we use the normalization method introduced in \cite{FO} to produce
a right moving frame. In planar centro-affine geometry, the group action is the linear
action of $\mathrm{SL}(2,\R)$ on $\R^2$. The following normalization conditions,
\begin{equation}\label{canor}
g \gamma = \begin{pmatrix} 1\\ 0\end{pmatrix}, \hskip 4ex g \gamma' = \begin{pmatrix} 0\\ \ast\end{pmatrix}
\end{equation}
uniquely determine the group element $g$. 
(In \eqref{canor}, the star denotes an arbitrary entry that plays no part in the
normalization; in fact, for the uniquely determined $g$, it is equal to $v$.)

The left moving frame is $\rho=g^{-1}=(\gamma \hskip 1ex\frac{1}{v} \gamma')$; 
a direct calculation shows that the Maurer-Cartan matrix is given by \eqref{MC}.
\qed
\end{pf}

The results in \cite{M1}  show that if a group of the form $G\semip\R^n$, with $G$ semisimple, acts
on $\R^n$ in an affine fashion, then, under certain conditions of arc-length
preservation, the space of periodic Maurer-Cartan matrices is a Poisson manifold.
The Hamiltonian structure is defined by first rewriting this space as
the quotient of $\Lo \g^\ast$ by the action of a subgroup of $\Lo G$. 
Then, the Poisson bracket (\ref{br1})
is Poisson reduced to that quotient, while (\ref{br2}) reduces only in some cases. 
For more information, we refer the reader to \cite{M1}.

In the centro-affine case the group action is not affine, but linear (a case for which geometric Poisson
structures have not been studied yet). Yet, for the case considered in this paper, we can explicitly compute the reduced Poisson brackets.

\begin{prop}\label{caqu}
The space of periodic Maurer-Cartan matrices of the form \eqref{MC},
in the neighborhood of a generic point,
can be identified with the quotient $\Lo \mathfrak{sl}^\ast(2)/\Lo N$,
where $N \subset SL(2,\R)$ is the subgroup of matrices of the form
$\left(\begin{smallmatrix} 1 & * \\ 0 & 1 \end{smallmatrix}\right)$,
and the action of $\Lo N$ on $\Lo\sl(2)^\ast$ is the gauge action 
\begin{equation}\label{gaugea}
g\cdot L = g^{-1} g' + g^{-1} L g, \qquad g\in \Lo N, \, L \in \Lo\sl(2)^*.
\end{equation}
\end{prop}
\begin{pf} 
Let $m \in \Lo \sl(2)$ (after identification of $\sl(2)$ with its dual). We can locally solve the system $\eta' = \eta m$ to obtain a fundamental matrix solution $\eta(x) \in \SL(2)$. This matrix will have a monodromy in the group, since $m$ has periodic coefficients. Let us denote by $\gamma$ the first column of $\eta$, so that $\gamma$ has the same monodromy as $\eta$. This will be the planar curve associated to $\eta$. Let $\rho$ be its moving frame, as in Proposition \ref{camf}. Given that the differential invariants of $\gamma$ are periodic, 
$\rho$ has the same monodromy as $\eta$.

Then, $\eta g=\rho$ for $g\in \mathcal{L}SL(2)$ given by
\[
g = \eta^{-1} \rho = \begin{pmatrix} \gamma &\ast \end{pmatrix}^{-1}\begin{pmatrix}\gamma& \frac1s \gamma'\end{pmatrix} = \begin{pmatrix} 1& \ast \\ 0&\ast\end{pmatrix}.
\]
This shows that $g\in \Lo N$. 
Moreover, the action $\eta \rightarrow \eta g$ on solutions of the equation $\eta'=\eta m$ induces the gauge action \eqref{gaugea} on $m$, taking $m$ to the Maurer-Cartan matrix of $\gamma$.
\qed
\end{pf}
\begin{thm} 
The Poisson brackets (\ref{br1}) and (\ref{br2}) 
can be Poisson reduced to the subspace 
$\mathcal{Z}\subset \Lo \sl(2)^\ast$ consisting of loops of the form
\begin{equation}\label{mcca}
\begin{pmatrix} 0&a\\ b&0\end{pmatrix}.
\end{equation}
The resulting Poisson brackets are, respectively,
\begin{equation}\label{rbr1}
\{h, f\}(a,b) = \int_{S^1} \begin{pmatrix}\displaystyle\frac{\delta f}{\delta a} &\displaystyle\frac{\delta f}{\delta b} \end{pmatrix}
\begin{pmatrix} -\frac12 D\frac1b D\frac1bD +\frac1b a D + D\frac 1b a&0\\0&0\end{pmatrix}  
\begin{pmatrix}\displaystyle\frac{\delta h}{\delta a} \\[8pt] \displaystyle\frac{\delta h}{\delta b} \end{pmatrix} dx
\end{equation}
and
\begin{equation}\label{rbr2}
\{h,f\}_0 = \{h, f\}(a,b) = \int_{S^1} \begin{pmatrix}\displaystyle\frac{\delta f}{\delta a} &\displaystyle\frac{\delta f}{\delta b} \end{pmatrix}
\begin{pmatrix} \frac2bD&0\\0&0\end{pmatrix}  \begin{pmatrix}\displaystyle\frac{\delta h}{\delta a} \\[8pt]\displaystyle\frac{\delta h}{\delta b} \end{pmatrix} dx,
\end{equation}
where $D = \frac d{dx}$and $h,f$ are functionals defined on $\mathcal{Z}$,
and we identify points in $\mathcal{Z}$ with pairs $(a,b)$ of periodic functions.

By further restricting the Poisson brackets to the subspace for which $b=1$, 
one obtains the Poisson operators  $J=-\frac12 D^3 + aD + Da$ and $J_0=2D$, 
which define the bi-Hamiltonian structure of the KdV equation.
\end{thm}
\begin{pf} 
We show how to calculate the reduced Poisson brackets, and refer the reader to \cite{M2} for more details. In Proposition \ref{caqu}, we showed that the space of matrices of the form \eqref{mcca} can be identified with the quotient $\Lo\sl(2)^\ast/\Lo N$. In order to reduce the brackets to this quotient, we consider two functionals $h, f$ defined on $\Lo\sl(2)^\ast/\Lo N$ and extend them to $\Lo \sl(2)^\ast$ requiring both extensions to be constant along the orbits the gauge action of $\Lo N$ on $\Lo\sl(2)^\ast$.
We denote by $H$ and $F$ the variational derivatives of the extended functionals evaluated at a section of the quotient space. Let $\n$ be the Lie algebra associated to $N$. Then the condition on $\mathcal{H}$ translates infinitesimally as
\[
H' + [K, H] \in \n^0
\]
for any $K$ with zero diagonal elements, where $\n^0$ is the annihilator of $\n$ (under the Killing form). 
Also, since $\Hop$ is an extension of
$h$, the form of its gradient must be
\[
H = \begin{pmatrix} \alpha & \frac{\delta h}{\delta b}\\ \frac{\delta h}{\delta a} & -\alpha\end{pmatrix}.
\]
Since $\n$ consists of strictly upper triangular matrices, we can identify $\n^0$ with the set of upper triangular matrices. Thus,
\[
H'+[K, H]= \begin{pmatrix} \alpha' + a \frac{\delta h}{\delta a} - b \frac{\delta h}{\delta b}&\left(\frac{\delta h}{\delta b}\right)' - 2a\alpha\\ \left(\frac{\delta h}{\delta a}\right)'+2b\alpha& - \alpha' - a \frac{\delta h}{\delta a} + b \frac{\delta h}{\delta b}\end{pmatrix} \in\n^0
\]
implies $\alpha = -\frac1{2b}  \left( \frac{\delta h}{\delta a}\right)'$, which completely determines $H$.
The same condition determines the variational derivative of $\mathcal{F}$.
Finally, the reduced Poisson brackets are obtained from formulas \eqref{br1}, and \eqref{br2}  applied to these extensions:
\[
\{h,f\}(a,b) = \int_{S^1} \tr\left((H'+[K,H])F\right) dx \]
\[=\int_{S^1} \frac1{2b}\left(\frac{\delta f}{\delta a}\right)'
\left(\frac1b \left(\frac{\delta h}{\delta a}\right)'  \right)' - 
\frac1b\left(\frac{\delta f}{\delta a}\right)' (a \frac{\delta h}{\delta a} - 
b\frac{\delta h}{\delta b}) + \left(\frac{\delta h}{\delta b}\right)' \frac{\delta f}{\delta a}
+\frac ab \frac{\delta f}{\delta a} \left(\frac{\delta h}{\delta a}\right)' dx,
\]
and
\[
\{h,f\}_0(a,b) = \int_{S^1} \tr\left(\left[ \begin{pmatrix} 0&1\\0&0\end{pmatrix}, H\right]F\right) dx
= \int_{S^1} 2\frac1b \left(\frac{\delta h}{\delta a}\right)' \frac{\delta f}{\delta a} dx,
\]
which coincide with (\ref{rbr1}) and (\ref{rbr2}).
\qed
\end{pf}
Notice that here we have used $\begin{pmatrix} 0&1\\0&0\end{pmatrix}$ instead of the projective
choice $\begin{pmatrix} 0&0\\1&0\end{pmatrix}$. This is due to the fact that $p$ appears in the
$(1,2)$  entry of the Maurer-Cartan matrix, while $\kappa$ appeared in the $(2,1)$ entry.

Applying the results of this section, and using the identification between 
the invariants of star-shaped curves and those projective curves,
induced by projectivization $\pi$ (described at the end of \S\ref{relatethem}),
we obtain the following 

\begin{thm} 
The projectivization map $\pi$ induces a local bi-Poisson map between the space of
periodic Maurer-Cartan matrices for centro-affine curves parametrized by centro-affine arc-length, and the
space of Maurer-Cartan matrices for parametrized projective curves.
\end{thm}

\section{KdV Potentials and the AKNS System}\label{KdVsection}
In this section, we will relate the moving frame for solutions of
the Schwarzian flow to the AKNS representation of KdV.
This relationship will be important for generating solutions of
Pinkall's flow.  We begin by pointing out a relationship between
the AKNS system and the projective moving frame.

Suppose that the evolution of the map $\phi:I \to \RP^1$ takes the
form
\begin{equation}\label{Schwarzshift}
u_t = (\kappa - a)u'
\end{equation}
for a constant $a$.  Then \eqref{kappatform} specializes to
$$\kappa_t = -\frac12 \kappa''' + 2\kappa \kappa' + (\kappa - a) \kappa'
= -\frac12 \kappa''' + 3 (\kappa - \tfrac{a}3)\kappa',
$$
showing that the function $q(x,t) = \kappa - \tfrac13 a$ satisfies the KdV equation:
\begin{equation}\label{ourKdV}
q_t = -\tfrac12 q''' + 3q q'.
\end{equation}

The AKNS system for the KdV equation \cite{AKNS, Fo}
takes the form $\psi' = U \psi, \psi_t = V \psi$ where
$\psi$ is a two-dimensional vector or matrix and
$$U = \begin{pmatrix} \lambda & q\\ 1 & -\lambda \end{pmatrix},
\qquad
V = \begin{pmatrix} -2\lambda^3 + \lambda q + \tfrac12 q' & -2\lambda^2 q
-\lambda q'  -\tfrac12 q'' +q^2 \\ q -2\lambda^2 & 2\lambda^3 - \lambda q -
\tfrac12 q'\end{pmatrix}.
$$
If we let $\varphi = \psi^T$, then we obtain the system
\begin{equation}\label{AKNS-KdV}
\dfrac{d\varphi}{d x}  = \varphi \begin{pmatrix}\lambda & 1 \\ q  & -\lambda
\end{pmatrix}, \qquad
\dfrac{d\varphi}{d t}  = \varphi \begin{pmatrix} -2\lambda^3 +\lambda q +\frac{1}{2} q' & q -2\lambda^2 \\
-\tfrac12 q'' -\lambda q' + q(q -2\lambda^2) & 2\lambda^3 -\lambda q-\frac{1}{2} q'
\end{pmatrix}.
\end{equation}
By comparing this with \eqref{rhoxform} and \eqref{rhotform}, we obtain
\begin{prop}
If $\phi$ satisfies the evolution equation \eqref{utrform} with $r = \kappa - 3\lambda^2$,
then the $\lambda$-normalized frame $\rho$ satisfies the AKNS system \eqref{AKNS-KdV}
with $q= \kappa - \lambda^2$.
\end{prop}
\begin{remark}
Since the KdV equation for $q$ is equivalent to the compatibility condition of system \eqref{AKNS-KdV}, the potential $q(x,t)=\kappa-\lambda^2$ must be assumed to be independent of the spectral parameter $\lambda$.
Thus, in order to construct a solution to the AKNS system
for a given KdV solution $q$ from the moving frame, we would have to  select $\lambda$-dependent initial conditions for the family of evolution equations with $r=\kappa-3\lambda^2$, so that the same potential $q$ results from all flows.
%  Namely, if we fix an initial condition $q(x,0) = q_0(x)$, then we will require that $\phi(x,0;\lambda)$ have curvature $\kappa$ equal to $q_0(x) + \lambda^2$.
\end{remark}
%
%From now on, we will consider only the evolution equation
%\begin{equation}\label{uequation}
%u_t = (\kappa - 3\lambda^2) u'
%\end{equation}
%for maps $\phi:I \to \RP^1$, inducing KdV potential
%$$q(x,t) = \kappa - \lambda^2.$$

We will now specialize to the Schwarzian KdV equation (i.e., setting $a=0$ in \eqref{Schwarzshift}),
and discuss a direct connection between this flow and the AKNS system, bypassing the moving frame.

\subsection{Closed Solutions of Pinkall's Flow}

As we will see, a star-shaped curve can be constructed in terms of the eigenfunctions
of the AKNS system \eqref{AKNS-KdV}. We will show that the constructed curve evolves by Pinkall's flow.

\begin{prop}\label{buildstar}
 Let $\Phi$ be a fundamental matrix solution of \eqref{AKNS-KdV}, satisfying the condition $\det \Phi =-1$. Define the curve $\gamma_0 \in \R^2$ as the second column of $\Phi$ evaluated at $\lambda=0$:
\begin{equation}\label{gammarecon}
\gamma_0=\left. \Phi \right|_{\lambda=0}  \begin{pmatrix} 0 \\ 1 \end{pmatrix}.
\end{equation}
Then $\gamma_0$ is a star-shaped curve with centro-affine curvature $p=-q$. Moreover $\gamma_0$ evolves by Pinkall's flow \eqref{pinkall}.
\end{prop}

\begin{pf}
We introduce
\begin{equation}\label{gammalambda}
\gamma=\Phi \begin{pmatrix} 0 \\ 1 \end{pmatrix},
\end{equation}
for arbitrary $\lambda$, and use the spatial part of the AKNS system \eqref{AKNS-KdV} to compute
\[
\gamma'=\Phi' \begin{pmatrix} 0 \\ 1 \end{pmatrix} = \Phi \begin{pmatrix} 1 \\ 0 \end{pmatrix}-\lambda \gamma.
\]
Then, $\det (\gamma, \gamma')=\det \left(\Phi \begin{pmatrix} 0 \\ 1 \end{pmatrix} , \Phi \begin{pmatrix} 1 \\ 0 \end{pmatrix} \right)=-\det{\Phi}=1$, and thus $\gamma$ is an arclength parametrized centro-affine curve for every $t$. Moreover,
\[
\gamma''=\Phi \begin{pmatrix} \lambda & 1 \\ q & -\lambda \end{pmatrix} \begin{pmatrix} 1 \\ 0 \end{pmatrix} -\lambda \gamma'=\lambda \Phi \begin{pmatrix} 1 \\ 0 \end{pmatrix} + q \gamma -\lambda \left[ \Phi \begin{pmatrix} 1 \\ 0 \end{pmatrix}-\lambda \gamma \right]=(q+\lambda^2) \gamma,
\]
showing that the centro-affine curvature of $\gamma$ is $-(q+\lambda^2).$

Finally, using the temporal part of the AKNS system \eqref{AKNS-KdV}, 
we compute the velocity field of $\gamma$:
\begin{equation}\label{gammarecont}
\begin{split}
\gamma_t & =\Phi \begin{pmatrix} q-2\lambda^2  \\ 2\lambda^3 -\lambda q -\frac12 q' \end{pmatrix} = \\
& = (q-2\lambda^2) (\gamma' +\lambda \gamma) +\left(2\lambda^3 -\lambda q -\frac{q'}{2}\right) \gamma=-\frac{q' }{2}\gamma + (q-2\lambda^2) \gamma'.
\end{split}
\end{equation}
Evaluating \eqref{gammarecont} at $\lambda=0$ gives Pinkall's flow for a curve $\gamma_0$ with curvature $p=-q.$
\qed
\end{pf}

If we construct the curve $\gamma$ at $\lambda\not=0$, 
we can use the following symmetry of the KdV equation and its AKNS system to obtain another solution of Pinkall's flow, with a different but related centro-affine curvature function:
\begin{prop}\label{AKNSsymmetry}
Suppose $\Phi$ satisfies the AKNS system \eqref{AKNS-KdV} at $(q,\lambda)$. Then
\begin{equation}\label{gauge}
\tilde{\Phi}(x,t)=\Phi(x+ct, t) \begin{pmatrix} 1 & 0 \\ a & 1\end{pmatrix},
\end{equation}
with $c=-3\lambda^2-6\lambda a,$ satisfies the AKNS system at $(\tilde{q}, \lambda+a)$, with
\[
\tilde{q}(x,t)=q(x+ct, t)-a^2-2\lambda a.
\]
\end{prop}
\begin{pf}
Although this symmetry is the well-known Galilean symmetry of the KdV equation and its Lax pair, we sketch the main steps of the proof in the AKNS setting. Let $\tilde{\Phi}=\Phi A$, with $A$ a gauge matrix possibly dependent on $x$; then
\[
\tilde{\Phi}'=\tilde{\Phi}\left[ A^{-1}\begin{pmatrix} \lambda & 1 \\ q & -\lambda \end{pmatrix} A + A^{-1}A'\right].
\]
It is easy to check that $A$ must be equal to  
$\begin{pmatrix} 1 & 0 \\ a & 1 \end{pmatrix}$ 
in order for the matrix expression in the square brackets to be of the form 
$\begin{pmatrix} \lambda+a & 1 \\ \tilde{q}  & -(\lambda + a) \end{pmatrix}$.
It follows that $\tilde{q}=q-a^2-2\lambda a.$ Finally,  using system \eqref{AKNS-KdV} to compute $\tilde{\Phi}_t=(c\Phi_x+\Phi_t)A$, one finds that choosing $c=-3a^2-6\lambda a$ will leave the form of the temporal part of the AKNS system invariant, with $\lambda$ replaced by $\lambda+a$ and $q$ replaced by $\tilde{q}.$
\qed
\end{pf}

By setting $a=-\lambda$ in the previous proposition, and applying Prop. \ref{buildstar},
it immediately follows that
\begin{cor}\label{shiftcor}
Let $\Phi$ be as in Proposition \ref{AKNSsymmetry} and let $\gamma$ be as defined by 
\eqref{gammalambda}. Then, for any $\lambda$, the curve
$\tilde{\gamma}(x,t)=\gamma(x+3\lambda^2t, t)$ solves equation \eqref{pinkall}
with $\tilde{p}(x,t)=p(x+3\lambda^2t, t)-\lambda^2.$
\end{cor}

The symmetry property derived in Proposition \ref{AKNSsymmetry} allows us to reconstruct closed solutions of Pinkall's flow given  periodic solutions of the KdV equation, by selecting  appropriate values of $\lambda$.

\begin{remark} The components of $\gamma$ solve the Schr\"odinger equation
\begin{equation}\label{schr}
\psi''+(\zeta-q)\psi=0, \qquad \zeta=-\lambda^2.
\end{equation}
In fact, e.g., $\gamma_1=\phi_2$ the second component of a solution of system \eqref{AKNS-KdV},  and a short calculation shows that $\phi_2$ satisfies \eqref{schr}.
\end{remark}
It follows that, for a given periodic $q$, only the values of  $\zeta$ for which the eigenfunction $\psi$ of the Schr\"odinger operator $\mathcal L = -(\partial_x)^2 + q$ is periodic give rise to a closed curve. We can finally state the following

\smallskip
\noindent
{\bf Closure Condition.} {\em Given a periodic KdV potential $q,$  the associated solution of Pinkall's flow $\gamma$ is a closed curve provided $\zeta=-\lambda^2$ is a periodic point of the spectrum of $q$.}

\section{Finite-gap solutions of Pinkall's flow}
In this section we will construct examples of curves evolving by Pinkall's flow,
associated with finite-gap solutions of the KdV equation.  These curves will be
built from linearly independent solutions $\psi_1, \psi_2$ of the scalar Lax pair (of which
\eqref{schr} is the spatial part).  First, we review the formulas for these solutions.
(The following is based on sections 3.1 and 3.4 in \cite{bbeim}, with suitable
changes in normalization.)

\subsection{Finite-Gap KdV Solutions}
We keep the same normalization for the KdV equation \eqref{ourKdV}; the complete
expression for its (scalar) Lax pair is
$$
\left\{
\begin{aligned}
\mathcal L \psi &= \zeta\psi, & \mathcal L &= -(\partial_x)^2 + q,\\
\psi_t &= \mathcal P \psi, & \mathcal P &= -2 (\partial_x)^3 + 3 q \partial_x + \tfrac32 q_x.
\end{aligned}\right.
$$
Let $\Sigma$ be a hyperelliptic Riemann surface of genus $g$ with real branch points,
described by
$$\mu^2 = \prod_{j=0}^{2g} (\zeta - E_j), \qquad
E_0 < E_1 < \ldots < E_{2g}.$$
(Note that $\zeta=\infty$ is also a branch point.)
Then there
are meromorphic differentials $\rd \Omega_1,\rd \Omega_2$ on $\Sigma$, with certain prescribed singularities at $\infty$, such that when $P \in \Sigma$ lies over $\zeta \in \C$,
a solution to the Lax pair is given by the {\em Baker-Akhiezer eigenfunction}
\begin{equation}\label{fgeigen}
\psi(x,t;P) = \exp(\Omega_1(P) x + \Omega_2(P) t)
\dfrac{\theta(\A(P) + V x + W t) \theta(D)}
{\theta(\A(P) + D)\theta(V x+ W t + D)}.
\end{equation}
Before giving the formula for the KdV solution, we will explain how to calculate the ingredients
in this formula:

\medskip
\noindent
1. Choose a homology basis on $\Sigma$, with cycle $a_k$ on
the upper sheet winding counterclockwise around the branch cut between
$E_{2k-1}$ and $E_{2k}$, and cycles $b_k$ such that the intersection
pairing satisfies $a_j \cdot b_k = \delta_{jk}$ and $b_j \cdot b_k = 0$,
where $1\le j,k \le g$ (see, e.g., Figure \ref{genus2spectrum}).
Let $\omega_k$ be the basis of holomorphic differentials on $\Sigma$
such that $\oint_{a_j} \omega_k = 2\pi \ri \delta_{jk}$.  Let $B_{jk} = \oint_{b_j} \omega_k$,
and define the Riemann theta function, used in \eqref{fgeigen}, as
$$\theta(z) = \sum_{n \in \mathbb Z^g} \exp\langle n, z + \tfrac12 B n\rangle,
\quad z \in \C^g.$$

\medskip\noindent
2.  Construct Abelian differentials of the form
$$\rd\Omega_1 = \dfrac{\ri}2 \dfrac{\zeta^g}{\mu} \rd\zeta + \ldots,\qquad
\rd\Omega_2 = 3 \ri \dfrac{(\zeta^{g+1}-\tfrac{c}2\zeta^g)}{\mu} \rd\zeta
 + \ldots,
$$
where $c=\sum_{j=0}^{2g}E_j$ and the missing terms are linear combinations
of the $\omega_k$ chosen uniquely so that $\oint_{a_k} \rd \Omega_1=0$ and $\oint_{a_k}\rd\Omega_2=0$.
The vectors $V$ and $W$ in \eqref{fgeigen} are the $b$-periods of these
differentials, i.e., $V_k = \oint_{b_k}\rd\Omega_1$ and $W_k = \oint_{b_k} \rd\Omega_2$.
Because of the reality of the branch points, these vectors are pure imaginary; the
vector $D$ is also pure imaginary, with arbitrary components.

\medskip\noindent
3. The Abelian integrals $\Omega_i(P)$ are computed with basepoint $E_0$, while
the Abel map $\A(P) = \int^P \omega$ (where $\omega$ is the vector of normalized
holomorphic differentials) is computed with basepoint $\infty$.  Both of these integrals
fail to be path-independent; however, we make \eqref{fgeigen} well-defined by
requiring that the integration paths in $\A(P)$ and $\Omega_i(P)$ differ
by a fixed path from $\infty$ to $E_0$ in $\Sigma_0$, the Riemann surface cut open
along the homology cycles.

\bigskip
The formula for $u(x,t)$ is derived by substituting \eqref{fgeigen} into the
spatial part of the Lax pair, and taking the limit as $\zeta \to \infty$.  One obtains
\begin{equation}\label{fgsolution}
q(x,t) = -2\left( c_1 + (\partial_x)^2 \log \theta(V x + W t + D)\right).
\end{equation}
The real constant $c_1$ is defined by the asymptotic expansion of $\Omega_1$ near $\infty$ in $\Sigma_0$:
$$\Omega_1 = w^{-1} - c_1 w + O(w^3),$$
where $w$ is a holomorphic coordinate near $\infty$ such that
$w^2 = -1/\zeta$ and
\[
\mu = \ri (-1)^{g-1} w^{-2g-1}(1+\tfrac{c}2 w^2 + O(w^4)).
\]

\begin{remark} The values of $V,W$ and $c_1$ can be obtained
by calculating the $a$-periods of the coordinate differentials $(\zeta^i/\mu)\rd\zeta$
(which are holomorphic on $\Sigma$ for $i <g$).
Namely, if constants $c_{jk}$ are such that $\omega_j = c_{jk} (\zeta^{g-k}/\mu) \rd\zeta$,
then
$$V_j = 2\ri c_{j1}, \qquad W_j =2(2\ri c_{j2}+\tfrac{c}2 V_j).$$
Furthermore,
$$c_1 = -\dfrac{c}2 - \dfrac{1}{4\pi}\sum_{k=1}^g V_k \oint_{a_k}\dfrac{\zeta^g}{\mu}\rd\zeta.$$
These formulas are obtained by applying Stokes' Theorem to certain differentials
on $\Sigma_0$; see (\cite{bbeim}, \S2.4.3) and (\cite{CI}, \S5.3) for similar calculations.
\end{remark}

\begin{remark}
The Riemann theta function is $2\pi \ri$-periodic in each entry.  Thus, $q(x,t)$ will have period $L$ in $x$ if  $\frac{L}{2\pi \ri}V$ is an integer vector.  (This will occur, for an appropriate choice of $L$, if the entries of $V$ are rationally related.)  Assuming this is the case, we can consider the Floquet spectrum of $q$.  The eigenfunction $\psi$ is $L$-periodic iff $L\Omega_1(P)$ is also an integer multiple of $2\pi \ri$.  In fact, taking the Baker eigenfunctions at points lying over the same value of $\zeta$ generically gives linearly independent solutions to the Lax pair, and we can use these to compute that the Floquet discriminant is given by
$$\Delta(\zeta) = 2 \cos(\ri L \Omega_1(P)).$$
\end{remark}

\subsection{Generating Examples}
In our examples, we choose real branch points so that the vector $V$ has rationally related components.
We then construct a curve in the centro-affine plane, evolving
by Pinkall's flow, by using shifted eigenfunctions as specified in Corollary \ref{shiftcor}.
We use Baker eigenfunctions evaluated at certain points $P$ (lying over the continuous spectrum of $q$) chosen so as to make the curve smoothly closed.
In particular, given a finite-gap KdV solution of period $L$ and a point $P$ at which
$$\dfrac{L\Omega_1(P)}{2\pi \ri} = \dfrac{m}{n}$$
for a rational number $m/n$ in lowest terms, then the eigenfunctions will
have period $nL$.  (In the examples below, $L=\pi$ in each case.)
Then the curve will close up smoothly after $n$ periods of its curvature, hence
will at each time be congruent to itself under an $SL(2)$-motion of period $n$.

\begin{figure}[hp] %  figure placement: here, top, bottom, or page
\centering
\includegraphics[width=5.5in]{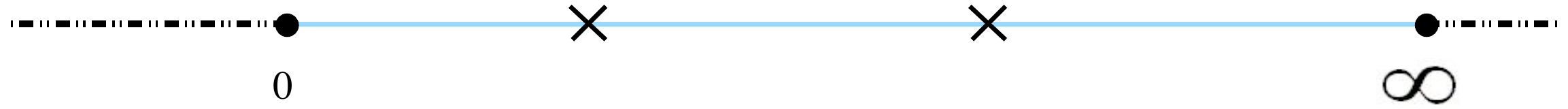} %\vskip .1in
\includegraphics[width=5.5in]{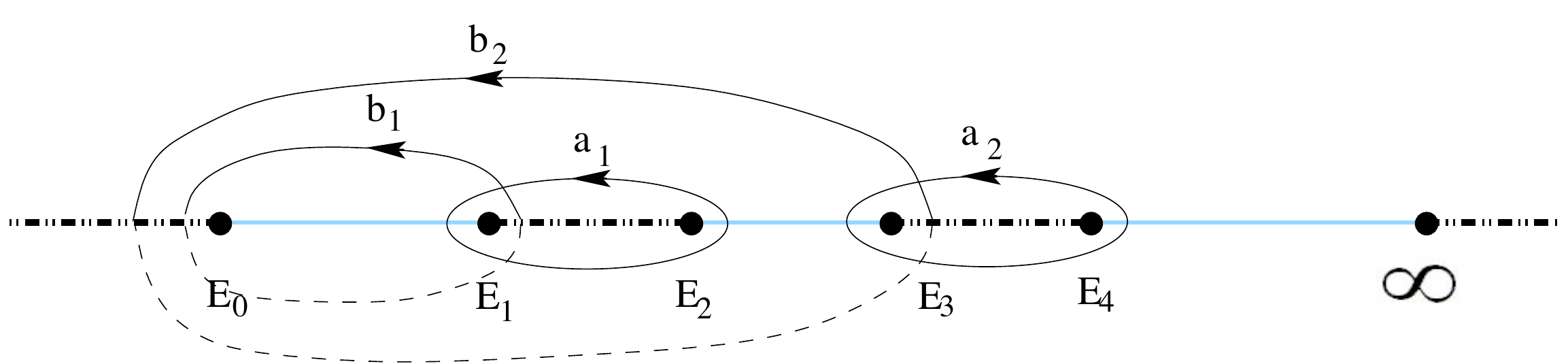}
\caption{At top, the spectrum of the zero KdV potential, with two marked double points to be opened up by isoperiodic deformation; at bottom, the spectrum of the resulting genus 2 potential.}
\label{genus2spectrum}
\end{figure}

We obtain suitable locations for the branch points and $P$ by carrying out isoperiodic deformations, beginning with the spectrum of the zero potential, and opening up successive gaps.  (The ODE system for isoperiodic deformations was derived in \cite{GS} and the scheme for deforming to higher genus was described for the NLS equation in \cite{CI2}. This carries over to the KdV case without significant modifications.)  In more detail, the continuous spectrum of the zero potential is $[0,\infty)$, and contains periodic points where $\zeta$ is the square of an integer.  We obtain genus 2 solutions of period $\pi$ by opening up gaps around $\zeta=4,9$ (see Figure \ref{genus2spectrum}), or genus 3 solutions by opening up gaps around $\zeta=1,4,9$.

\begin{figure}[hp] %  figure placement: here, top, bottom, or page
\centering
\includegraphics[width=2.2in]{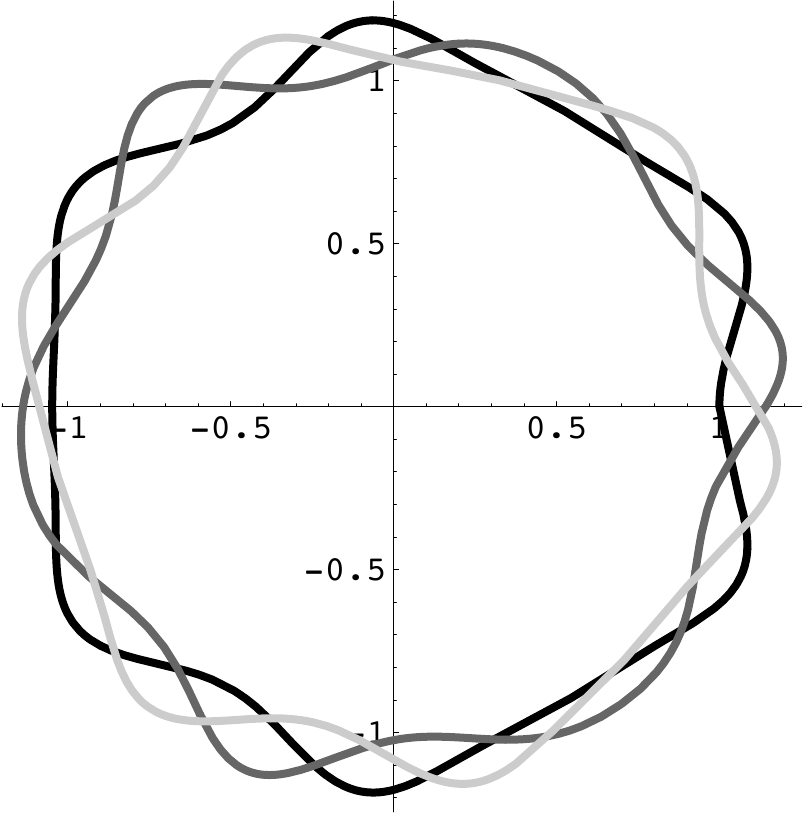} \quad
\includegraphics[width=1.7in]{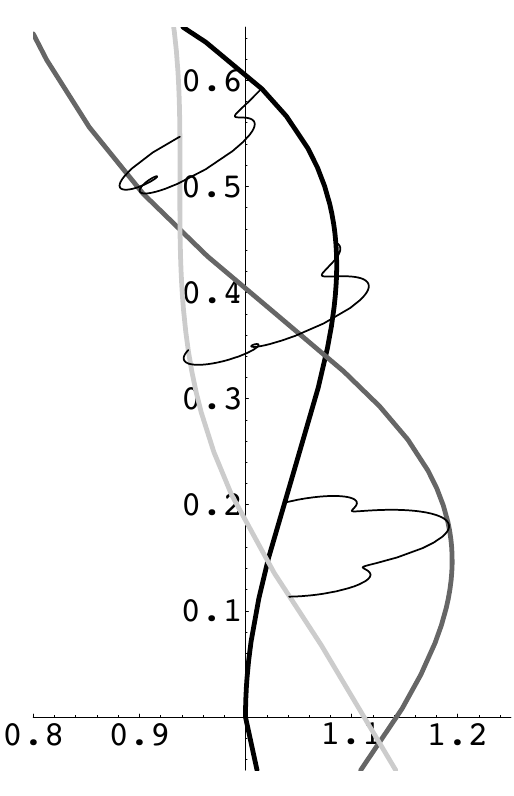}
\caption{Solution generated by a genus 2 potential, with branch points
$(-0.0400525, 3.47762, 4.47468, 8.62255, 9.42258)$,
frequency vector $V=(-4\ri, -6\ri)$ and $D=0$,
constructed at $\zeta = 0.400062$; the value $\Omega_1=(2/3) \ri$ gives 3-fold symmetry.  The first picture shows the curve at times $0, .08, .16$ (darker to lighter); second picture shows time evolution of three particles along the curve.}
\label{genus2simple}
\end{figure}

\begin{figure}[h] %  figure placement: here, top, bottom, or page
\centering
\includegraphics[width=1.7in]{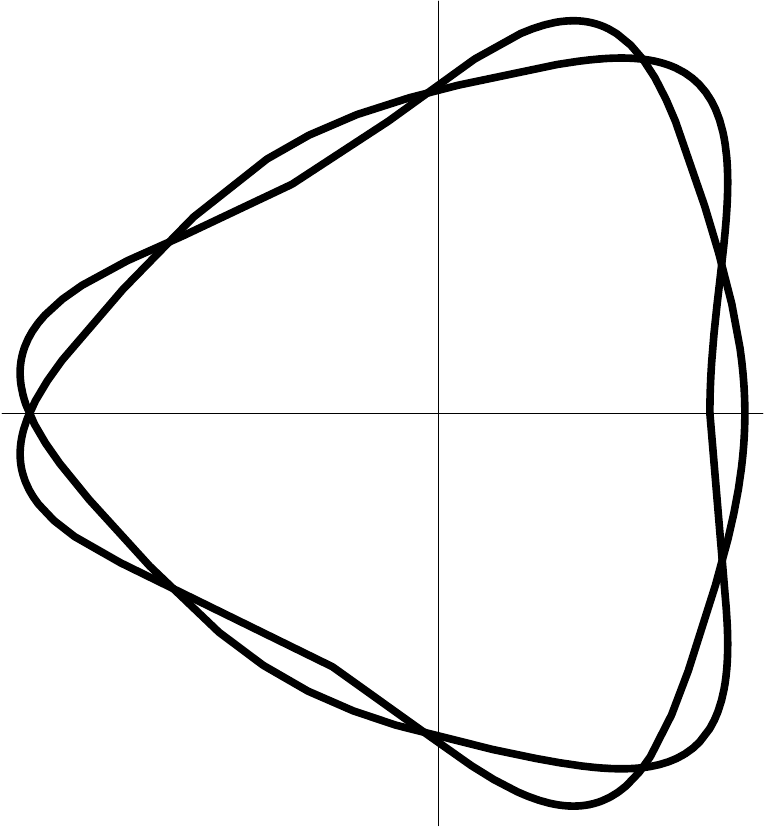}
\includegraphics[width=1.8in]{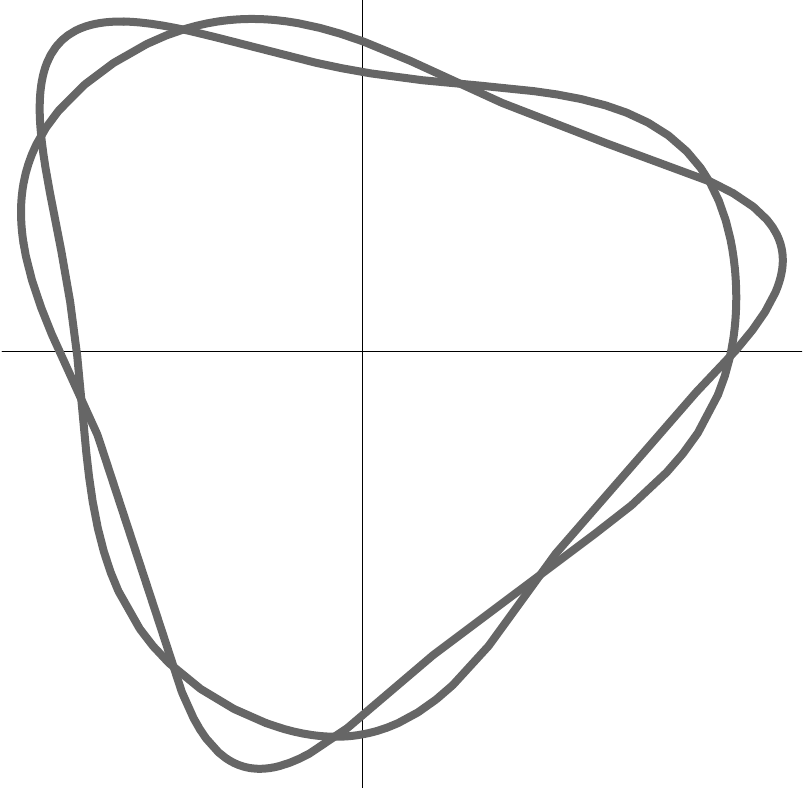}
\includegraphics[width=1.85in]{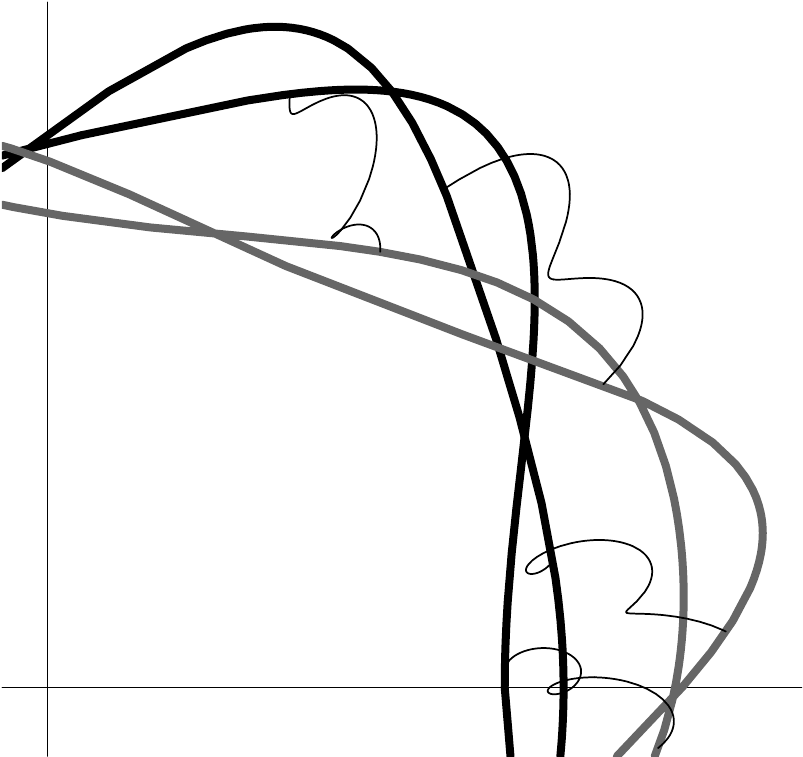}
\caption{Solution generated by a genus 2 potential, with branch points
$(-0.12408, 3.11086$, $4.68657$, $8.15068, 9.95102)$,
frequency vector $V=(-4\ri, -6\ri)$ and $D=0$, constructed at $\zeta=1.58191$, where $\Omega_1=(4/3)\ri$. Shown at times $0$ and $1.35$ (dark and light, respectively),
along with the time evolution of 4 particles along the curve in the first quadrant.}
\label{genus2double}
\end{figure}

\begin{figure}[h] %  figure placement: here, top, bottom, or page
\centering
\includegraphics[width=1.8in]{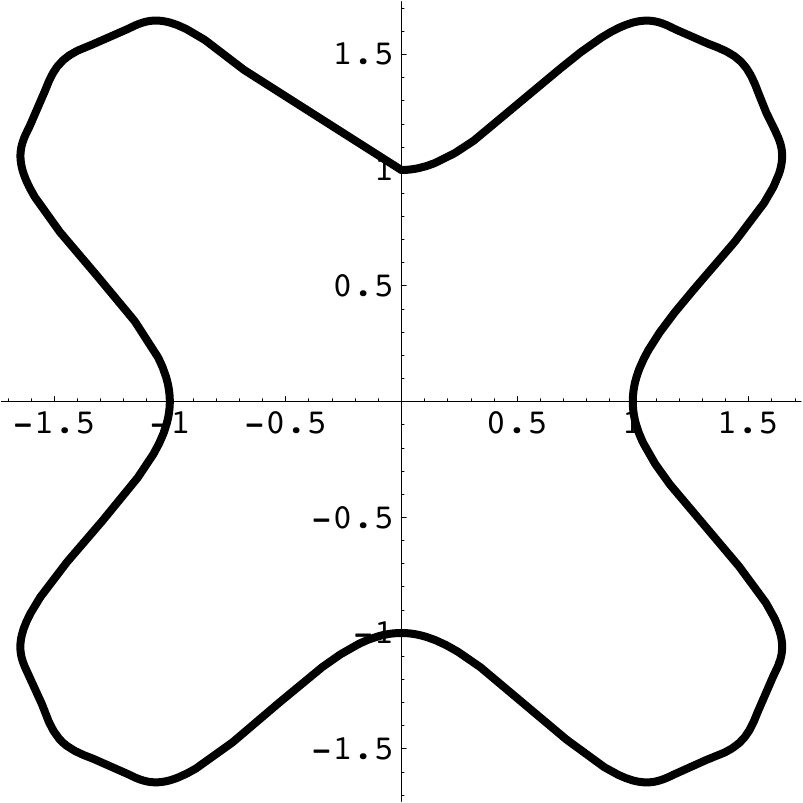}
\includegraphics[width=2in]{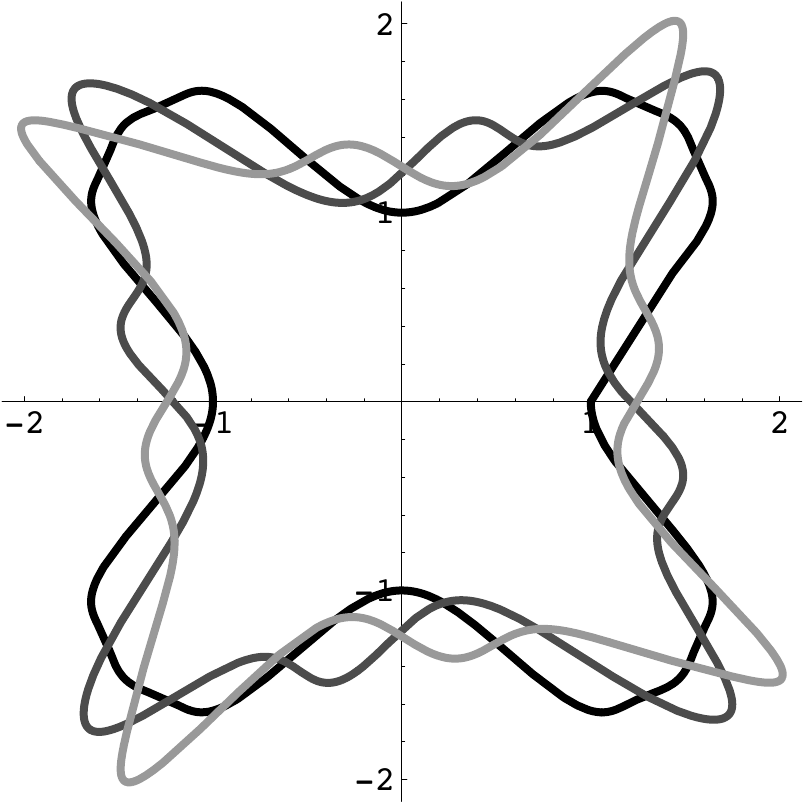}
\includegraphics[width=1.5in]{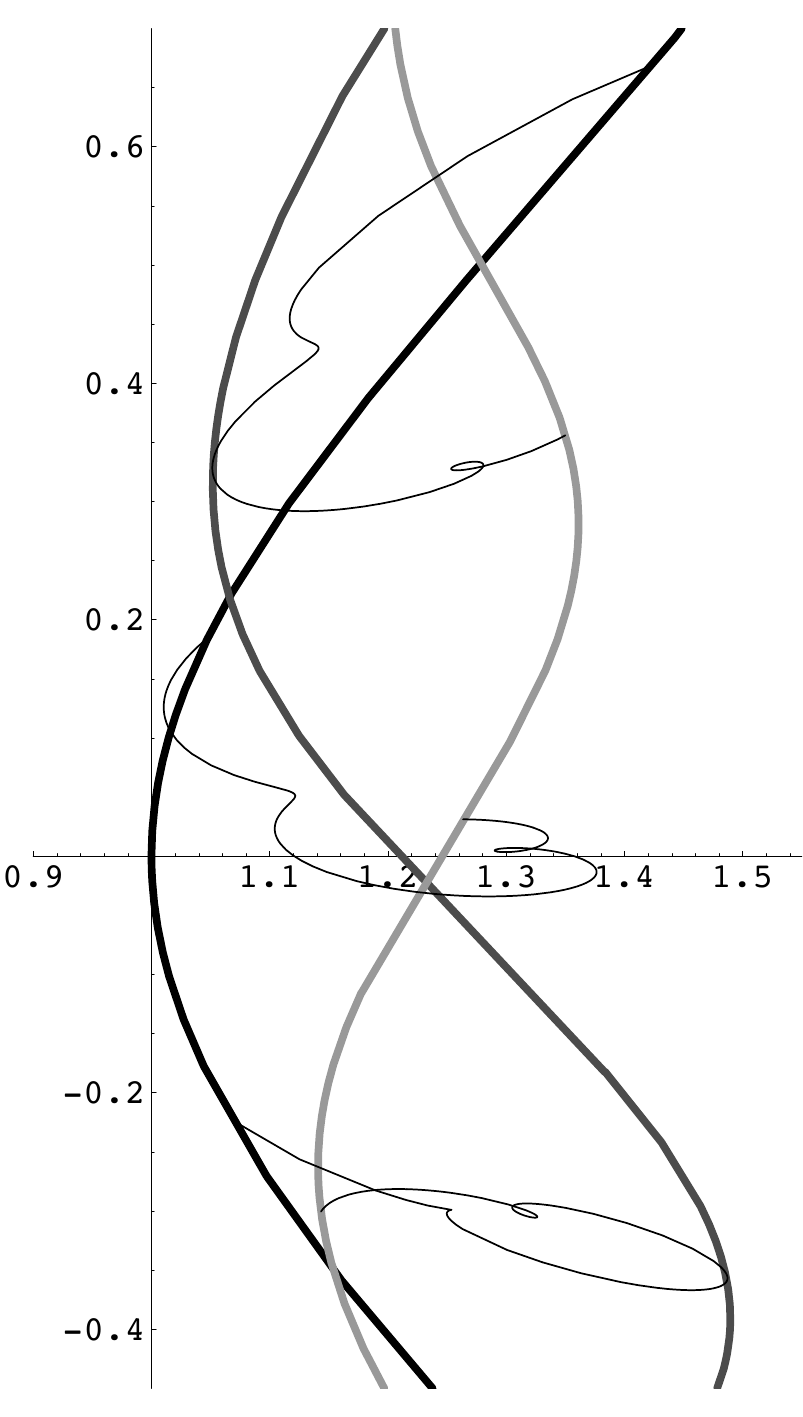}
\caption{
Solution generated by a genus 3 potential, with branch points
$(-0.277121$, $0.291382$, $1.23879$, $2.95364$, $4.93069$, $8.30355$, $9.90382)$,
frequency vector $V=(-2\ri, -4\ri, -6\ri)$ and $D=0$, constructed at $\zeta = -0.0705768$; the value $\Omega_1= -.5 \ri$ gives 4-fold symmetry.
The first picture shows the initial condition, followed by curves at times $0, 0.08, .16$ (from dark to light), then tracks of three particles under the time evolution.}
\label{genus3simple}
\end{figure}

\begin{figure}[h] %  figure placement: here, top, bottom, or page
\centering
\includegraphics[width=6in]{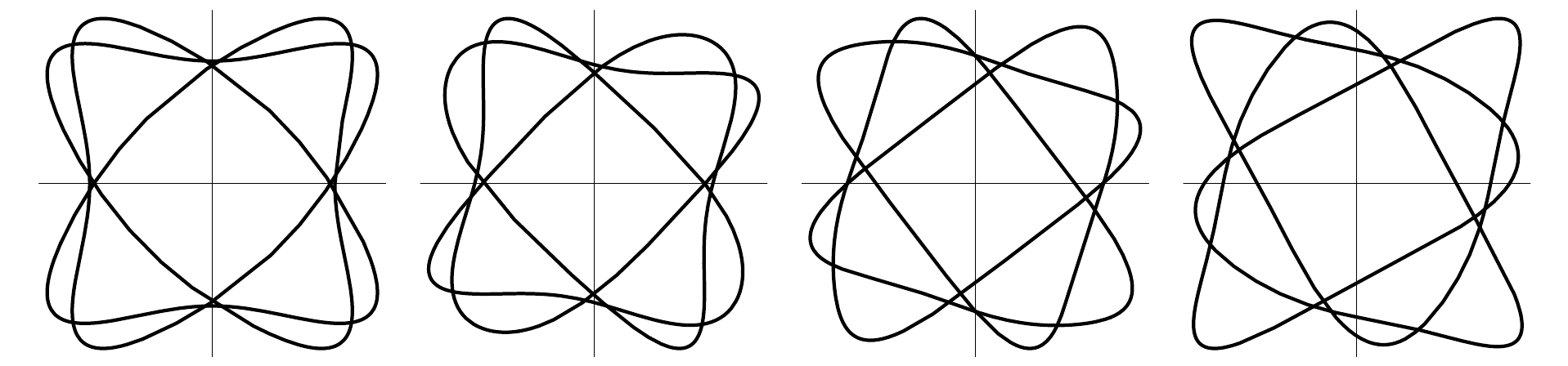}
\caption{A genus 3 solution, with the same branch points and $V,D$ values as in Figure \ref{genus3simple}, but now the double point is $\zeta=2.01811$, where $\Omega_1 =1.5\ri$.
The curve is shown at times $0, .02, .04, .06$.}
\label{genus3triple}
\end{figure}

The captions to Figures \ref{genus2simple} through \ref{genus3triple} give
further details for our examples.

\section{Acknowledgements}
A.~Calini and T.~Ivey gratefully acknowledge support of this work by the National Science Foundation under grant number DMS-0608587.


\begin{thebibliography}{9}
\newcommand{\ditto}{{\leavevmode\vrule height 2pt depth -1.6pt width 23pt\,}}

\bibitem{AKNS} M.J. Ablowitz, D.J. Kaup, A.C. Newell, H. Segur, {\em The inverse scattering transform-Fourier analysis for nonlinear problems},  Stud. Appl. Math. 53, pp 249-315, 1974.

\bibitem{bbeim}
{Belokolos~E., Bobenko~A., Enolskii A.~B.~V., Its~A. and Matveev~V.}
 {\em {A}lgebro-{G}eometric {A}pproach to {N}onlinear {I}ntegrable {E}quations},
  Springer, 1994.

\bibitem{CQ1} {K-S. Chou~and C.~Qu.}
{\em The KdV Equation and Motion of Plane Curves}, J. Phys. Soc. Japan, {\bf 70} (7), pp 1912-1916, 2001.

\bibitem{CQ2}  {K-S. Chou~and C.~Qu.}
{\em Integrable motions of space curves in affine geometry}, Chaos, Solitons and Fractals, {\bf 14}, pp 29-44, 2002.

\bibitem{CI}
{A.~Calini and T.~Ivey}
{\em Finite-Gap Solutions of the Vortex Filament Equation: Genus One Solutions and Symmetric Solutions},  J. Nonlinear Sci., {\bf 15}, pp. 321--361, 2005.

\bibitem{CI2}
\ditto, {\em Finite-Gap Solutions of the Vortex Filament Equation:  Isoperiodic Deformations},
J. Nonlinear Sci., {\bf 17}, pp. 527--567, 2007.

\bibitem{DS} V.~G. Drinfel'd and V.~V. Sokolov.
\newblock Lie algebras and equations of {K}orteweg-- de {V}ries type,
\newblock In {\em Current problems in mathematics, Vol. 24}, Itogi Nauki i
  Tekhniki, pages 81--180. Akad. Nauk SSSR Vsesoyuz. Inst. Nauchn. i Tekhn.
  Inform., Moscow, 1984.


\bibitem{FO} M. Fels, P.J. Olver, {\em Moving coframes. II. Regularization and theoretical foundations}, Acta Appl. Math., pp. 127--208, 1999.

\bibitem{Fo} A.P. Fordy, {\em A Historical Introduction to
Solitons and B\"acklund Transformations}, in ``Harmonic Maps and Integrable
Systems'' (ed. Fordy and Wood), Vieweg, 1994.

\bibitem{GS} {P.G.~Grinevich and M.U.~Schmidt}, {\em Period preserving nonisospectral flows and the moduli space of periodic solutions of soliton equations: the nonlinear Schr\"{o}dinger equaton}, Physica D {\bf 87}, pp 73-98,  1995.

%\bibitem{HS} R. Huang, D.A. Singer, {\em A new flow on starlike curves in $\R^3$}, Proceedings  of the AMS, 130-9, pp 2725-2735, 2002.

\bibitem{H}
E. Hubert.
\newblock Generation properties of differential invariants in the moving frame methods,
\newblock{preprint}.

\bibitem{KN} I.M. Krichever and S.P. Novikov, {\em Holomorphic bundles over algebraic curves and nonlinear equations}, Russ. Math. Surv. {\bf 35} (6), pp 53Ð79, 1980.

\bibitem{LHM} A. Lopez,  R. Heredero, G. Mar\'\i~Beffa, {\em Invariant Differential Equations and the Adler-Gel'fand-Dikii bracket} , J. Math. Phy. {\bf 38}, pp 5720--5738, 1997.

\bibitem{M3} G. Mar\'\i~Beffa, {\em The Theory of Differential Invariants and KdV Hamiltonian Evolutions}, Bull. Soc. Math. France, {\bf 127} (3), pp 363-391, 1999.

\bibitem{M1} G. Mar\'\i~Beffa, {\em Poisson geometry of differential invariants of curves in some nonsemisimple homogenous spaces}, Proc. Amer. Math. Soc. {\bf 134}, pp 779-791, 2006.

\bibitem{M2} G. Mar\'\i~Beffa, {\em Hamiltonian structures on the space of differential invariants of curves in flat semisimple homogeneous spaces}, submitted.

%\bibitem{Oc}
%T. Ochiai, {\em Geometry associated with semisimple flat homogeneous spaces},
%Transactions of the AMS, 152,  pp 159--193, 1970.

%\bibitem{O} P.J. Olver {\em  Equivalence, Invariance and Symmetry}, Cambridge University Press, Cambridge, 1995.

\bibitem{P95} U. Pinkall, {\em Hamiltonian flows on the space of star-shaped curves},
Result. Math. {\bf 27}, pp 328--332, 1995.

\bibitem{We} J. Weiss, {\em The Painlev\'e property for partial differential equations. II: B\"{a}cklund transformation, Lax pairs, and the Schwarzian derivative},  J. Math. Phys. {\bf 24}, 1405, 1983.

%\bibitem{Wi} E.J. Wilczynski, {\em Projective differential geometry of curves and ruled surfaces},
% B.G. Teubner, Leipzig, 1906.

\end{thebibliography}
\end{document}